\documentclass[aps,floatfix,prd,twocolumn,a4paper,10pt,citeautoscript,
preprintnumbers,superscriptaddress,showpacs]{revtex4-1}

\usepackage[english]{babel}	

\usepackage{amsmath}		
\usepackage{amssymb}		
\usepackage{bbm}		    
\usepackage{bm}			

\usepackage{graphicx}			
\usepackage{graphics}			
\DeclareGraphicsExtensions{.pdf}	

\usepackage[colorlinks=true, pdfstartview=FitV,linkcolor=blue,
			citecolor=blue, urlcolor=blue]{hyperref}

\newcommand{\bs}{\boldsymbol}
\newcommand{\Eqref}[1]{\eqref{#1}}				
\newcommand{\Figref}[1]{Fig.~\ref{#1}}			
\newcommand{\Exp}[1]{\mathrm{e}^{#1}}

\newcommand{\Grad}{\bs \nabla}

\newcommand{\Curl}{\bs \nabla\times}

\newcommand{\A}{{\bs A}}
\newcommand{\B}{{\bs B}}
\newcommand{\D}{{\bs D}}
\newcommand{\F}{\mathcal{F}}
\newcommand{\Hc}[1]{\mathrm{H}_{c#1}} 
\newcommand{\Q}{\mathcal{Q}}
\newcommand{\Dq}[1]{\Delta_{\mbox{\scriptsize $\Q_{#1}$}}}
\newcommand{\Dmq}[1]{\Delta_{\mbox{\scriptsize $-\Q_{#1}$}}}

\newcommand{\Int}{{\mbox{\tiny Int}}}
\newcommand{\PDW}{{\mbox{\tiny PDW}}}
\newcommand{\dW}{{\mbox{\tiny $d$-wave}}}
\newcommand{\Dpdw}{\Delta_\PDW}
\newcommand{\Ddw}{\Delta_d}
\newcommand{\qh}{{\hat q}}

\begin{document}
\title{Checkerboard order in vortex cores from pair-density-wave superconductivity}
\author{Daniel~F.~Agterberg}
\affiliation{Department of Physics,
University of Wisconsin-Milwaukee, Milwaukee, WI 53211}
\author{Julien~Garaud}
\affiliation{Department of Theoretical Physics,
KTH-Royal Institute of Technology, Stockholm, SE-10691 Sweden}
\date{\today}

\begin{abstract}
 We consider competing pair-density-wave (PDW) and $d$-wave superconducting states in a magnetic field. We show that PDW order appears in the cores of $d$-wave vortices, driving checkerboard charge-density-wave (CDW) order in the vortex cores, which is consistent with experimental observations. Furthermore, we find an additional CDW order that appears on a ring outside the vortex cores. This CDW order varies with a period that is twice that of the checkerboard CDW and it only appears where both PDW and $d$-wave order co-exist. The observation of this additional CDW order would provide strong evidence for PDW order in the pseudogap phase of the cuprates. We further argue that the CDW seen by nuclear magnetic resonance at high fields is due to a PDW state that emerges when a magnetic field is applied.

\end{abstract}
\maketitle

\section{Introduction}

Pair-density-wave (PDW) superconducting order has emerged as a realistic candidate for order in the charge-ordered region of the pseudogap phase of the cuprates near one-eighth filling. It naturally accounts for both superconducting (SC) correlations and for static quasi-long-range charge-density-wave (CDW) order observed near this hole doping and at temperatures below approximately 150 K \cite{ghi12,com14,sil14,wu14,agt08,ber09,lee14}, and it can explain observed signatures of broken time-reversal symmetry \cite{sid13,xia08,he11,kam02,kar14,Agterberg.Melchert.ea:15}. Moreover, PDW can lead to the quantum oscillations seen in the cuprates \cite{zel12} and can also explain anomalous quasi-particle properties observed by angle-resolved photoemission (ARPES) measurements \cite{lee14}.  In addition, numerical simulations of theories of a doped Mott insulator reveal PDW order to be a competitive ground-state to $d$-wave superconductivity \cite{cor14}. It is therefore important to find experiments that can identify PDW order in the cuprates. Motivated by the observation of checkerboard CDW order inside $d$-wave vortex cores by scanning tunneling microscopy (STM) \cite{hof02,lev05} and by nuclear magnetic resonance (NMR) \cite{wu11,wu13}, we examine the competition between $d$-wave and PDW superconductivity in applied magnetic fields.  Previous theoretical studies of competing orders in a magnetic field have emphasized competing spin-density-wave (SDW) order \cite{zha02,sac04}, CDW order \cite{zha02,sac04,ein14}, and staggered flux phases \cite{lee01,web06} with $d$-wave superconductivity. Competing PDW and $d$-wave order has not been extensively studied (note that superconducting phase disordered PDW competing with $d$-wave order has been examined \cite{che04}). Here, we find that inside the vortex cores of $d$-wave superconductivity, PDW order drives the observed checkerboard CDW order and, in conjunction with  $d$-wave superconductivity, it also drives an additional CDW order that appears in a ring-like region outside the vortex cores. This additional CDW order has twice the period of the observed checkerboard CDW order and serves as a smoking gun for PDW order.

\begin{figure}[!htb]
\hbox to \linewidth{ \hss
\includegraphics[width=0.95\linewidth]{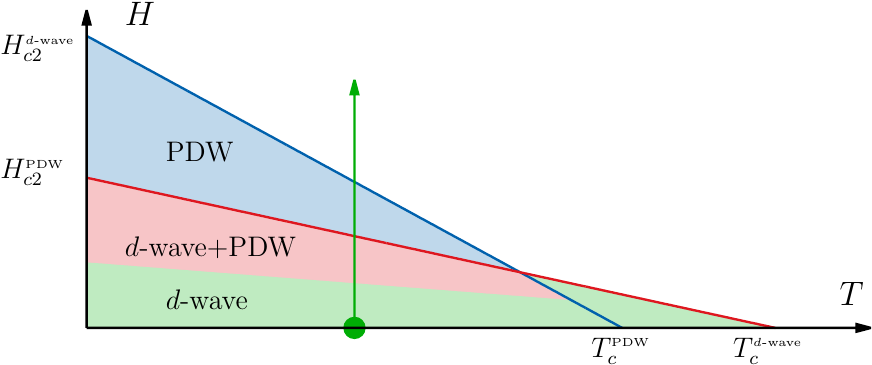}
\hss}
\vspace{-0.2cm}
\caption{
(Color online) --
Sketch of the field/temperature phase diagram of the model with
competing order. For low applied fields, the $d$-wave ($\Ddw$)
is present, and it completely suppresses the PDW ($\Dpdw$).
This is the green (lower) region of the phase diagram.
When increasing the external field, the $d$-wave order is
substantially suppressed, eventually triggering a phase
transition where the PDW overcomes the competition with
the $d$-wave and develops a nonzero averaged density.
For sufficiently low temperatures, the second critical field
of the PDW order exceeds that of the $d$-wave order. As a result,
further increase of the external field completely suppresses
the $d$-wave order, leaving only $\Dpdw$, which survives up to
$H=\Hc{2}^\PDW$, as shown in the blue (upper) region of the diagram.
The PDW order qualitatively accounts for the emergence of CDW at high
fields, provided the superconducting order of the PDW is suppressed
by phase fluctuations.
}\label{Fig:HTdiagram}
\end{figure}

In the following, we develop a phenomenological theory for competing PDW and $d$-wave superconductivity, sketched in \Figref{Fig:HTdiagram}. We assume that in zero field, only $d$-wave superconductivity appears at the expense of the PDW order. The PDW order can only appear when the $d$-wave order is weakened by the external field. This is followed by an analysis of the core structure of a single $d$-wave vortex, where we show that PDW order appears inside these cores, without any phase winding, generating the CDW order discussed above. Finally, we examine the behavior of this competing system as the field is further increased and identify a transition at which PDW order develops phase coherence and forms a vortex phase. At the mean-field level, PDW order simultaneously breaks gauge invariance and translational symmetry. Fluctuations can lead to two separate transitions: one for which gauge symmetry is broken and one for which translational symmetry is broken \cite{ber09-2}. We argue that at high fields, the superconducting order is removed by phase fluctuations, leaving behind the CDW order seen through NMR experiments.
%
\section{Ginzburg Landau theory of competing \texorpdfstring{$d$}{d}-wave and PDW superconductivity}

To investigate the physics resulting from the $H$-$T$ phase
diagram shown in \Figref{Fig:HTdiagram}, we consider a model
with competing $d$-wave and PDW superconductivity. The PDW
order parameter is represented by a four component complex
vector $\Dpdw$, defined as
$\Dpdw^\dagger=(\Dq{x}^*,\Dmq{x}^*,\Dq{y}^*,\Dmq{y}^*)$
and the $d$-wave by one complex (scalar) field $\Ddw$.
For an external applied field ${\bs H}$, which we will
take to be along the $z$-axis, ${\bs H}=H{\bs e}_z$,
the Ginzburg-Landau free-energy density is
\begin{equation}
\label{FreeEnergy}
\F=\frac{\B^2}{2}-\B\cdot{\bs H}+\F_\dW+\F_\PDW+\F_\Int\,,
\end{equation}
where $\B=\Curl\A$ is the magnetic field and $\A$ its vector
potential. $\F_\PDW$ describes the pair-density-wave $\Dpdw$,
and $\F_\Int$ its coupling to the $d$-wave order that obeys
\begin{equation}
\label{FreeEnergyDWave}
\F_\dW= \frac{1}{2}|\D\Ddw|^2
+\alpha_d|\Ddw|^2+\frac{\beta_d}{2}|\Ddw|^4	\,,
\end{equation}
with $\D=\Grad+ie\A$. Symmetry arguments dictate that the
free energy of the PDW has the following structure \cite{agt08}:
\begin{align}
\label{FreeEnergyPDW}
\F_\PDW &=
\frac{1}{2}\sum_{\qh,j}k_{\qh,j}|D_j \Delta_\qh|^2
+\sum_\qh\left(\alpha+\frac{\beta}{2}|\Delta_\qh|^2\right)|\Delta_\qh|^2
\nonumber\\
&+\gamma_1\left(|\Dq{x}|^2|\Dmq{x}|^2+|\Dq{y}|^2|\Dmq{y}|^2	\right)
\nonumber\\
&+\gamma_2\left(|\Dq{x}|^2+|\Dmq{x}|^2\right)
		  \left(|\Dq{y}|^2+|\Dmq{y}|^2\right)
\nonumber \\
&+\frac{\gamma_3}{2}\left(
\Dq{x}^*\Dmq{x}^*\Dq{y}\Dmq{y}+c.c.
\right)	\,.
\end{align}
Here, we neglect variations along the $z$-axis, thus $j=x,y$ is
the spatial index, while $\qh$ is a wave-vector index:
$\qh=\Q_x,-\Q_x,\Q_y,-\Q_y$. In the following, another convenient
index $q=\Q_x,\Q_y$, will also be used. The coefficients
$k_{\qh,j}$ of the kinetic term satisfy the following relation
$k_{\pm\Q_{x},x}=k_{\pm\Q_{y},y}\equiv1-k$ and
$k_{\pm\Q_{x},y}=k_{\pm\Q_{y},x}\equiv1+k$, and $k$ measures
the anisotropy of the system
\footnote{
Although it is physically relevant, the anisotropy $k$
has very little influence on the physics we describe here.
We verified that indeed for $k\neq0$, moderate anisotropies
do not qualitatively change the physical properties we discuss.
Thus in the rest of the paper, we consider only the isotropic
case $k=0$.
}.
Here $\Q_x$ represents the wavevector ${\bs\Q}_x=(\Q,0)$, $\Q_y$ represents ${\bs\Q}_y=(0,\Q)$, and $\Dq{x}$ represents the gap associated with the pairing between the fermion states $|{\bs k}+{\bs\Q}_x,\uparrow\rangle$ and $|{\bs k},\downarrow\rangle$, where ${\bs k}$ is the momentum and $\uparrow$, $\downarrow$ denote the spin-states. Our choice of the wavevectors and model for the PDW order is motivated by the recent proposal of Amperean pairing by P.A. Lee \cite{lee14}, for which it has been shown that PDW order can account for both the anomalous quasi-particle properties observed by ARPES and the CDW order (at momenta 2${\bs\Q}_x$ and 2${\bs\Q}_y$) observed in the pseudogap phase of Bi$_2$Sr$_{2-x}$La$_x$CuO$_{6+\delta}$ (Bi2201). Depending on the parameters $\gamma_i$, the free energy of the PDW sector \Eqref{FreeEnergyPDW} allows five possible distinct ground-states \cite{agt08}. We choose parameters such that, in the non-competing case, the PDW ground-state has the form $\Dpdw^\dagger=\Delta_0^*(1,1,i,i)$. This PDW ground-state is the same as that proposed in Ref.~\onlinecite{lee14} and is also found to be a ground-state in the spin-fermion model \cite{wan14,wan14-2}.

Both $\Ddw$ and $\Dpdw$ interact with the magnetic field
(through the kinetic terms) and are therefore indirectly
coupled. They also directly interact through $\F_\Int$:
\begin{align}
\label{FreeEnergyCoupling}
\F_\Int&= \gamma_4|\Delta_d|^2
\left(|\Dq{x}|^2+|\Dmq{x}|^2+|\Dq{y}|^2+|\Dmq{y}|^2\right)
\nonumber\\
&+\frac{\gamma_5}{2}\left(
\left[\Dq{x}^*\Dmq{x}^*+ \Dq{y}^*\Dmq{y}^*\right]\Ddw^2
+c.c.\right)\,.
\end{align}
The first term in \Eqref{FreeEnergyCoupling} is a bi-quadratic
coupling between the $d$-wave and the pair-density-wave
$\sim\gamma_4|\Delta_d|^2|\Delta_\PDW|^2$. The coexistence of
both order parameters is penalized for positive values $\gamma_4$,
and when strong enough, only one of the condensates supports a
nonzero ground-state density. Our choice of parameters is such
that when $H=0$, $\Ddw$ has lower condensation energy and $\Dpdw$
is completely suppressed, because of the interaction terms
\Eqref{FreeEnergyCoupling}. Moreover, as CDW order emerges
at high field, we require $\Dpdw$ to have a higher second 
critical field ($\Hc{2}^\PDW$) than $\Ddw$ ($\Hc{2}^\dW$).
These conditions lead to \Figref{Fig:HTdiagram}.
We note that in principle, the existence of the competing
PDW order can allow for the PDW driven CDW order to appear
in zero field in the vicinity of inhomogeneities or due to
fluctuations in some materials. Indeed CDW order has been
observed in YBa$_2$Cu$_3$O$_{6.67}$ in zero field through
high-energy x-ray diffraction \cite{cha12} (this CDW order
is enhanced by magnetic fields).

\section{PDW-driven CDW-order}

We take CDW order to be denoted by $\rho({\bs r})=\sum_\qh
\Exp{i{\bs\qh}\cdot{\bs r}}\rho_\qh$ (note that $\rho_{-q}=\rho_q^*$).
The coupling between $\rho_{2q}$ (with $q=\Q_x,\Q_y$) and PDW order
is given by \cite{agt08,ber09,lee14}:
\begin{equation}
\label{Rho2Qcoupling}
\sum_{q=\Q_x,\Q_y}\alpha_2|\rho_{2q}|^2
+\epsilon_2\left(\rho_{2q}\Delta_{-q}\Delta^*_{q}
+\rho_{-2q}\Delta_{q}\Delta^*_{-q}\right)	\,.
\end{equation}
Assuming that the CDW order is induced by the PDW order, we find that
\begin{equation}
\label{Rho2Q}
\rho_{\pm 2q}=\rho_{\mp 2q}^*=-\frac{\epsilon_2}{\alpha_2}
\Delta_{\pm q}\Delta^*_{\mp q} \,.
\end{equation}
The CDW order given by $\rho_{2q}$ corresponds to that observed in the pseudogap phase in zero field and to the checkerboard order observed inside the $d$-wave vortex cores. An important feature of this work is that the interplay between $d$-wave and PDW orders gives rise to an additional contribution to the CDW order. In particular, this coupling is given by \cite{agt08,ber09,lee14}
\begin{align}
\label{RhoQcoupling}
\sum_{q=\Q_x,\Q_y}\alpha_1|\rho_{q}|^2
+\epsilon_1 &\Big(
	\rho_{q}[\Delta_{-q}\Delta_d^*+\Delta_{q}^*\Delta_d]
\nonumber\\
&+\rho_{-q}[\Delta_{q}\Delta_d^*+\Delta_{-q}^*\Delta_d]
\Big)	\,.
\end{align}
Differentiation with respect to $\rho_{q}^*$ and $\rho_{q}$ yields the relations (this also assumes the CDW order is purely induced):
\begin{equation}
\label{RhoQ}
\rho_{\pm q}=\rho_{\mp q}^*=-\frac{\epsilon_1}{\alpha_1}
\left(\Delta_{\pm q}\Delta_d^*+\Delta_d\Delta_{\mp q}^*\right)
\,.
\end{equation}

The contributions $\rho_{\Q}$ and $\rho_{2\Q}$ to the CDW
are reconstructed according to
\begin{equation}
\label{CDWnQ}
\rho_{n\Q}=\sum_{q=\Q_x,\Q_y} \rho_{nq}\Exp{in{\bs q}\cdot{\bs r}}
+\rho_{-nq}\Exp{-in{\bs q}\cdot{\bs r}}	\,,
\end{equation}
which shows the $n$th-order contribution to the CDW. The CDW order
$\rho_{\Q}$ has twice the periodicity of $\rho_{2\Q}$ and is not
an induced order of the pure $\Dpdw$: it only appears
when both $\Ddw$ and $\Dpdw$ coexist. Consequently,  $\rho_{\Q}$
is a signature of the appearance of $\Dpdw$ in a $d$-wave
superconductor. Note that the existence of $\rho_{\Q}$ requires 
superconducting phase coherence for both the PDW and $d$-wave 
orders (strictly speaking, coherence in the phase difference 
between these two orders will suffice).
We note that an observation of $\rho_{\Q}$ has been reported 
\cite{bey09}, and below we make predictions about the
structure of $\rho_{\Q}$ around a vortex in $\Ddw$.

\section{Vortex properties and checkerboard pattern}

In order to investigate the interplay of $\Dpdw$ and $\Ddw$,
within the framework sketched in \Figref{Fig:HTdiagram},
we numerically minimize the free energy \Eqref{FreeEnergy}
both for single vortices and for a finite sample in external
field. The theory is discretized within a finite element formulation
\cite{Hecht:12} and minimized using a nonlinear conjugate gradient
algorithm (for detailed discussion on the numerical methods, see, for
example, \cite{Agterberg.Babaev.ea:14}).
\begin{figure}[htb]
\hbox to \linewidth{ \hss
\includegraphics[width=0.9\linewidth]{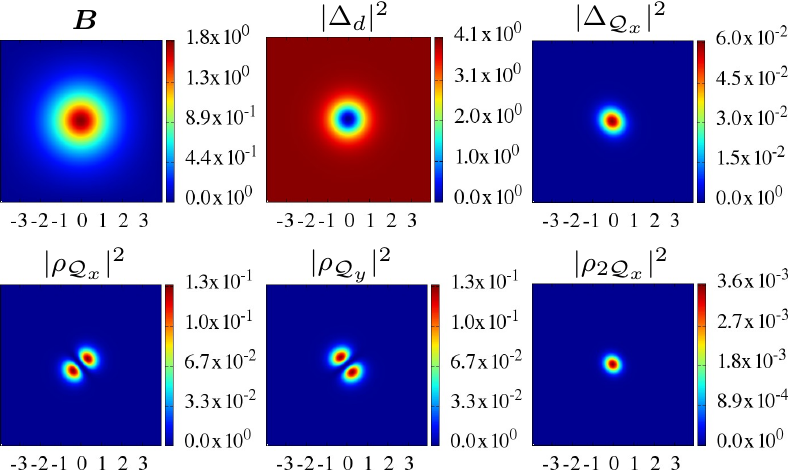}
\hss}
\vspace{-0.3cm}
\caption{
(Color online) --
The core structure of a single $d$-wave vortex. The parameters are
$(\alpha,\beta)=(-5,10)$ and $\gamma_2=\gamma_1/2=10\gamma_5=3$,
while the parameters for the $d$-wave order are
$(\alpha_d,\beta_d)=(-2.5,0.61)$. The parameters of the
interaction \Eqref{FreeEnergyCoupling} that directly couples
the PDW and the $d$-wave order are $\gamma_4=2$, $\gamma_5=0.5$ and
the gauge coupling constant is $e=0.4$.
The $d$-wave order has nonzero ground-state density and has a vortex,
while the components of the PDW are zero in the ground-state. At the
core of the $\Ddw$ vortex, because there is less density, it is 
beneficial for the components $\Ddw$ of the PDW to condense, as shown 
in the right panel of the first line (here we show only $\Dq{x}$ as 
the other components behave similarly). The second line displays the
induced CDW: $\rho_{2q}$ \Eqref{Rho2Q} and $\rho_q$ \Eqref{RhoQ}
(note $\rho_{2\Q_y}$ is similar to $\rho_{2\Q_x}$).
}
\label{Fig:Vortex}
\end{figure}

\begin{figure}[!b]
\hbox to \linewidth{ \hss
\includegraphics[width=0.95\linewidth]{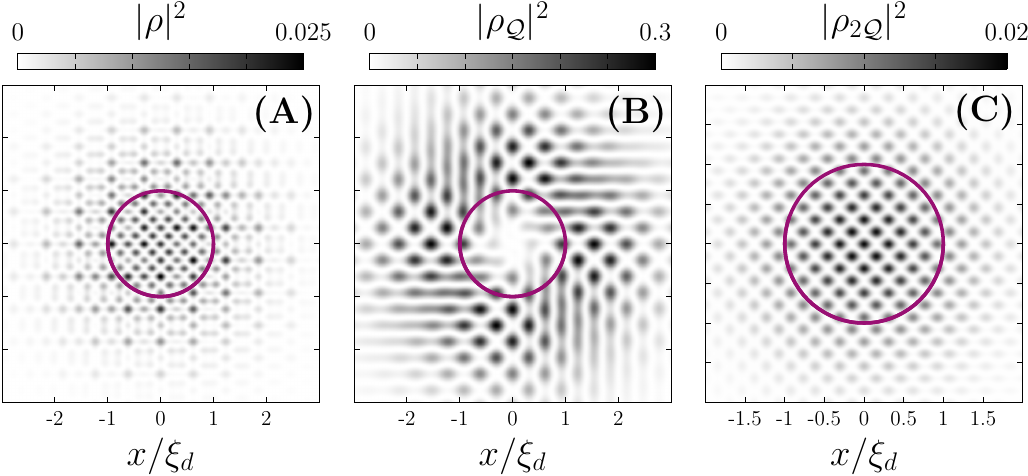}
\hss}
\vspace{-0.3cm}
\caption{
(Color online) --
The charge-density-wave order (a) and the contributions of
$\rho_{\Q}$ (a) and $\rho_{2\Q}$ (a), as defined in \Eqref{CDWnQ}.
The parameters are the same as in \Figref{Fig:Vortex} except
that $\gamma_5=1.0$. The circle of radius $\xi_d$, the coherence
length of the $d$-wave, indicates the size of the vortex core.
$\rho_{\Q}$ and $\rho_{2\Q}$ are shown for unit value of the ratios
$\epsilon_1/\alpha_1$ and $\epsilon_2/\alpha_2$, while $\rho$, the
total charge density, is shown for $\epsilon_1/\alpha_1=1$ and
$\epsilon_2/\alpha_2=0.1$. As a result, $\rho$ shows a checkerboard
in the vortex core. Furthermore, since $\rho_{\Q}$ varies with twice
the wavelength as $\rho_{2\Q}$, away from the core, every other peak
in $\rho$ is magnified.
}
\label{Fig:Checkerboard}
\end{figure}

Typical single vortex solutions (see \Figref{Fig:Vortex}) clearly
show that the components of the PDW order acquire small, yet nonzero
density at the center of the $d$-wave vortex core. As a result,
the CDW order is also nonzero at the vortex core. Far from the
vortex, the $\Dpdw$ decays to zero, and the induced CDW is suppressed
as well.
\Figref{Fig:Checkerboard} shows the magnitude of the total CDW
order as well as the contributions from different orders in $\Q$.
Here, we used the values $\Q=\pi/d$ and $d=4a_0$, where $a_0$
is the Cu-Cu distance in cuprates and, in qualitative accordance 
with experimental data \cite{fis07}, we take the $d$-wave coherence 
length to be $\xi_d=13a_0$. $\rho_{2\Q}$ forms a checkerboard 
pattern that extends significantly outside the vortex core, 
and this is consistent with the observations.

In addition to this checkerboard order, we also find that
$\rho_{\Q}$, which varies at twice the wave-length of $\rho_{2\Q}$,
is nonzero and also has a non-trivial structure. More precisely,
at the singularity in the $d$-wave, $\rho_{\Q}=0$, and when
$\Delta_d$ becomes nonzero, $\rho_{\Q}$ also becomes nonzero.
Since $\Dpdw$ exhibits no phase winding, $\rho_\Q$ inherits
the phase winding of $\Ddw$. A phase winding in $\rho_{\Q}$
implies a dislocation in the corresponding real-space order
\cite{Chaikin.Lubensky}. Consequently, the CDW order associated
with $\rho_{\Q}$ has a dislocation at the vortex core.
Since $\rho_{\Q}$ is suppressed in vortex cores, the checkerboard
pattern that appears there, is essentially due to $\rho_{2\Q}$.
The contribution of $\rho_{\Q}$ to the CDW becomes important
at distances larger than $\xi_d$. Moreover, as it varies with
a doubled wave-length, every other charge peak is magnified
in a region outside the core.
Note that away from the vortex, $\rho_{\Q}$ is suppressed
at a much slower rate than $\rho_{2\Q}$. Furthermore, if $\rho_{\Q}$
is observable at all, then it should vanish at $\Hc{2}^\dW$, while
$\rho_{2\Q}$, will persist to much higher fields.

\section{Field induced PDW and CDW orders}

\begin{figure}
\hbox to \linewidth{ \hss
\includegraphics[width=0.9\linewidth]{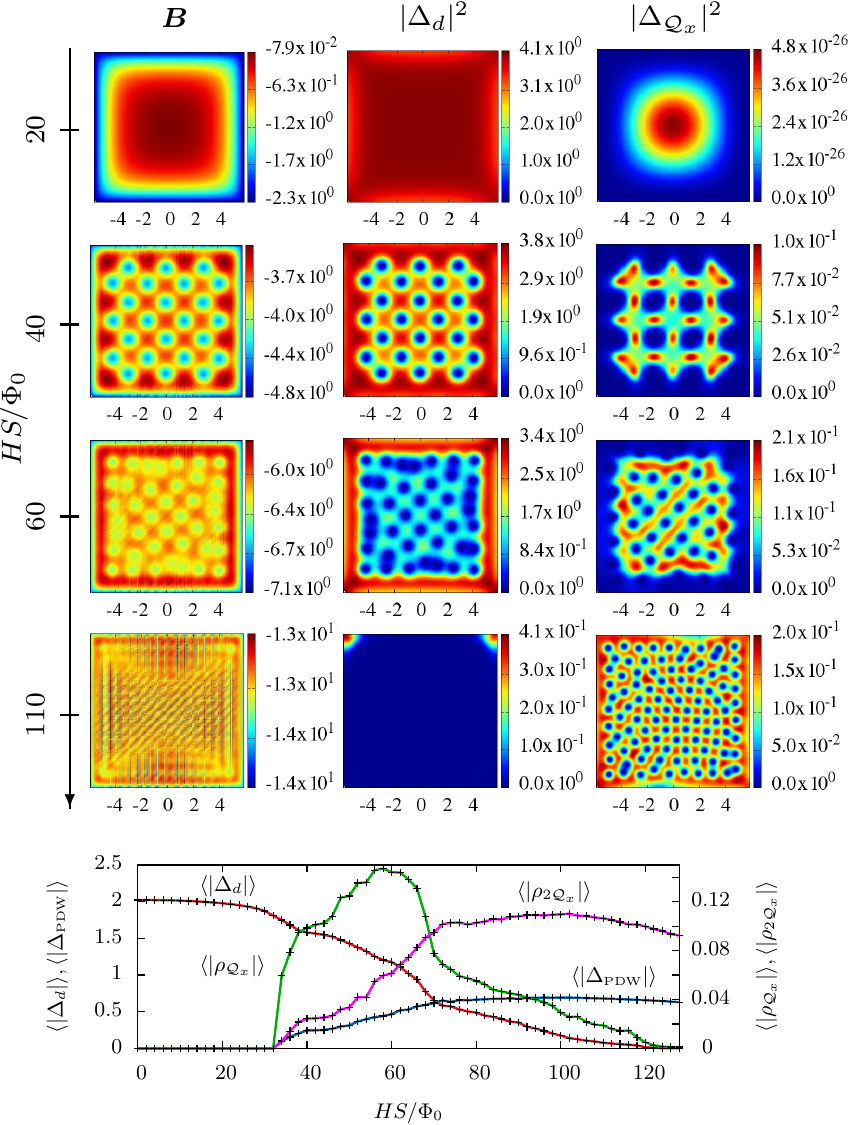}
\hss}
\vspace{-0.3cm}
\caption{
(Color online) --
Simulation over a finite sample with increasing values of the
external field (values are shown on the left) for the same
parameters as in \Figref{Fig:Checkerboard}.
The first column shows the magnetic flux, the second column shows
$|\Delta_d|^2$ and the last column shows $|\Dq{x}|^2$ (other
components of the PDW behave similarly to $\Dq{x}$).
The graph at the bottom shows order parameters averaged over
the sample, as functions of the applied field. There, we show
the densities of the $d$-wave and PDW order, as well as the
induced CDW contributions $\rho_{\Q_x}$ and $\rho_{2\Q_x}$.
Above a certain external field (here $HS/\Phi_0=32$), because
of the suppression of the $d$-wave order, the PDW develops a
nonzero expectation. The appearance of the PDW in external field
is also accompanied by induced CDW order.
}
\label{Fig:Magnetization}
\end{figure}

To investigate the evolution of the PDW and $d$-wave orders
in external field ${\bs H}$, for parameters corresponding to
\Figref{Fig:HTdiagram}, we minimize the free energy
\Eqref{FreeEnergy}, while imposing $\Curl\A={\bs H}$ at the
(insulating) boundary of the domain. We follow the vertical line
sketched in \Figref{Fig:HTdiagram}. That is, starting from $H=0$,
the field is sequentially increased after the solution for
the current value of $H$ is found. Typical results illustrating such
a simulation are shown in \Figref{Fig:Magnetization}.
In low fields, only $\Ddw$ has a nonzero ground-state
density and, as a result of the competition with $\Ddw$
in the interacting terms \Eqref{FreeEnergyCoupling},
$\Dpdw$ is fully suppressed (or vanishingly small).

Above the first critical field, vortices in $\Ddw$, carrying
a small amount $\Dpdw$ in their core, start entering the system.
The averaged PDW over the whole sample $\langle|\Dpdw|\rangle$
is still vanishingly small. With increasing field, the density
of vortices increases and they start to overlap
\footnote{
Note that vortices here arrange as squares. In principle anisotropies
or interactions originating in the complicated core structures
can account for this. Here we believe this is merely a finite
size effect. Indeed, the role of Meissner currents cannot be
neglected and definite statements about the lattice structures
cannot be safely made.
}
. That is, $|\Dpdw|$ and $|\Ddw|$ do not have ``enough room" to
recover their ground-state values. At this point, the lumps of
$\Dpdw$, previously isolated in vortex cores, interconnect and
$\Dpdw$ acquires a phase coherence globally. This behavior was
also found to occur in a similar system with competing orders
\cite{Garaud.Babaev:15}.
At this phase transition, not only does $\langle|\Dpdw|\rangle$
become nonzero, but the induced CDW $\rho_{\pm q}$ and $\rho_{\pm 2q}$,
also become nonzero on average (see \Figref{Fig:Magnetization}).
We conjecture that this phase transition is related to that seen
though NMR \cite{leb13}.

When the PDW order is on average nonzero, energetic considerations
dictate that it should acquire phase winding as well. Indeed, when
two condensates have nonzero density, the energy of configurations
that has winding in only one condensate diverges (at least logarithmically)
with the system size. As a result vortices in $\Delta_\qh$ are created
when $\langle|\Dpdw|\rangle\neq0$ \cite{Garaud.Babaev:15}.
Note that as it is still beneficial to have nonzero $\Dpdw$ inside
the vortex cores of $\Ddw$, the singularities that are formed due to
the winding in $\Delta_\qh$ do not overlap with those of $\Ddw$
[and they do not overlap with each other due to the terms $\gamma_i$
in \Eqref{FreeEnergyPDW}, which favor core splitting]. Thus, the CDW
order still appears within the vortex cores of $\Ddw$. Since all the
vortices that are created do not overlap with each other,
the magnetic induction is smeared out and is much more spatially
uniform than in usual vortex phases.

For fields above the second critical field of $\Ddw$,
only the PDW order survives. As a result, the contribution
$\rho_\Q$ to the induced CDW also vanishes and the observed
CDW order above $\Hc{2}^\dW$ is solely that induced by the
PDW (that is $\rho_{2\Q}$).
In this state, at the mean-field level, the vortices in $\Delta_\qh$ do not overlap, as the terms with $\gamma_i$ in \Eqref{FreeEnergyPDW} favor vortex core splitting. In principle, the parameters $\gamma_i$ can also be chosen so that the $\Delta_\qh$ cores coincide for some or all PDW components. This will not change the qualitative physics associated with the competition between $\Ddw$ and $\Dpdw$. However, it will affect the resulting high-field regime. In either case, we expect superconducting phase fluctuations to play an important role in the high-field phase. In particular, it is known that for type-II superconductors, high magnetic fields significantly enhance the role of fluctuations \cite{lee72,bla94}. Phase fluctuations will remove the superconducting long-range order of the PDW state, but the CDW order can still survive \cite{ber09-2}. A related mechanism was also considered in a different but related model of superconductivity \cite{Babaev:04a}.
%

\section{Conclusions}

We have considered a model of competing pair-density-wave
and $d$-wave superconductivity. The superconducting state
in the Meissner phase is purely $d$-wave. With increasing
external field, vortices in the $d$-wave superconductor are
formed and they carry PDW and induced CDW order in their core.
When these vortices significantly interact, the lumps of PDW
order acquire global phase coherence and both PDW and $d$-wave
superconductivity coexist. In the regions where both PDW and
$d$-wave order exist, the induced CDW order features a $\rho_\Q$
contribution that exists at twice the periodicity of the CDW
order observed in the pseudogap phase at zero fields. The observation
of $\rho_\Q$ can serve to identify the existence of PDW order
in the pseudogap phase.

\begin{acknowledgments}
We thank Egor Babaev, Andrey Chubukov,  Marc-Henri Julien, Manoj Kashyap,
Patrick Lee, and Yuxuan Wang for fruitful discussions.
DFA acknowledges support from NSF grant No. DMR-1335215.
JG was supported by  National Science Foundation under
the CAREER Award DMR-0955902 and by the Swedish Research
Council grants 642-2013-7837, 325-2009-7664.
The computations were performed on resources provided by the
Swedish National Infrastructure for Computing (SNIC) at the
National Supercomputer Center at Link\"oping, Sweden.
\end{acknowledgments}

%

\end{document}